# The Machine Learning Canvas: Empirical Findings on Why Strategy Matters More Than AI Code Generation

Martin Prause

**Abstract** Despite the growing popularity of AI coding assistants, over 80% of machine learning (ML) projects fail to deliver real business value. This study creates and tests a Machine Learning Canvas, a practical framework that combines business strategy, software engineering, and data science in order to determine the factors that lead to the success of ML projects. We surveyed 150 data scientists and analyzed their responses using statistical modeling. We identified four key success factors: Strategy (clear goals and planning), Process (how work gets done), Ecosystem (tools and infrastructure), and Support (organizational backing and resources). Our results show that these factors are interconnected - each one affects the next. For instance, strong organizational support results in a clearer strategy ($\beta = 0.432$, $p < 0.001$), which improves work processes ($\beta = 0.428$, $p < 0.001$) and builds better infrastructure ($\beta = 0.547$, $p < 0.001$). Together, these elements determine whether a project succeeds. The surprising finding? Although AI assistants make coding faster, they don't guarantee project success. AI assists with the "how" of coding but cannot replace the "why" and "what" of strategic thinking.

## I. INTRODUCTION

Empirical studies suggest that large language models (LLMs) can boost developer productivity by automating certain coding tasks. For example, developers using Copilot finish tasks up to 55% faster and experience a lighter cognitive workload [1]. This is particularly beneficial for less experienced programmers. Anthropic's analysis of approximately 500,000 coding sessions revealed that approximately 79% of interactions with Claude Code involve automation [2]. An empirical study demonstrated increased productivity in Python programming tasks; however, the researchers cautioned that effectiveness depends on task complexity and user expertise [3]. Furthermore, a large-scale field experiment conducted by the Bank for International Settlements revealed a 55% increase in productivity, with LLMs generating approximately one-third of the total lines of code [4]. While these findings affirm productivity improvements, they also imply an ongoing need for human oversight. Code generated by artificial intelligence (AI) addresses repetitive or well-defined programming activities rather than complex architectural decisions [5].

Although LLMs can accelerate code generation and reduce the cognitive burden of routine programming tasks, these micro-level efficiency gains do not translate into macro-level project success. In a study of 65 experienced data scientists, it was found that over 80% of AI projects fail — twice the rate of traditional IT projects [6]. The authors identified five primary failure modes: (1) misalignment between technical objectives and business problems, (2) poor data quality and infrastructure, (3) attempting to solve problems that are beyond the current capabilities of AI, (4) insufficient organizational readiness, and (5) inadequate governance structures. These findings confirm earlier research indicating high failure rates in machine learning (ML) deployments [7][8].

In response to these challenges, machine learning operations (MLOps) have emerged as a discipline that combines software engineering principles with machine learning-specific requirements [9]. A systematic literature review identifies MLOps as encompassing continuous integration and deployment, automated testing and validation, model versioning and governance, and production monitoring and maintenance [10]. Similarly, in another study the authors indicate that adoption success rates vary substantially depending on MLOps maturity. More mature firms demonstrate significantly better capabilities in terms of data management, automated deployments, and continuous integration and delivery [11].

These multifaceted challenges highlight the necessity of structured frameworks for planning, communicating, and executing AI and ML projects. The Business Model Canvas (BMC) approach, popularized by Osterwalder, has proven

Martin Prause, martin.prause@xinblue.de
.

effective in creating shared mental models among diverse stakeholders [12]. Recent applications to AI contexts include the AI Model Canvas, which adapts Canvas principles to the specific requirements of machine learning [13]. However, existing Canvas frameworks often fail to address the full complexity of ML projects. They typically focus on either technical aspects such as data pipelines and model architecture, or business considerations, such as value propositions and cost structures, without effectively integrating both perspectives. Additionally, these frameworks inadequately address the dynamic nature of ML projects, which require flexible yet structured approaches to iterative experimentation [14].

## II. Background: The Machine Learning Canvas

These limitations are addressed by developing a Machine Learning Canvas (MLC), which integrates organizational theory, software engineering, and data science. An empirical study using a Structural Equation Modeling (SEM) approach was conducted to identify the success determinants of ML projects in development contexts where large language models serve as coding assistants.

A business model is a blueprint that aligns a company's strategic objectives with its operational execution [15]. It explains how an organization creates, delivers, and captures value [16]. The hierarchical taxonomy identifies four fundamental dimensions of business models:
1. Value Proposition: What value is created, and for whom?
2. Value Architecture: How is value created through organizational capabilities?
3. Value Network: The ecosystem of relationships that enable value creation.
4. Value Finance: The economic model for capturing created value.

The Business Model Canvas operationalizes these dimensions through nine interconnected building blocks arranged on a visual canvas. This approach facilitates iterative development and enables stakeholders to maintain a holistic view while addressing specific components. ML projects share structural similarities with business models in that they focus on creating value from data by aligning multiple perspectives: business strategy, technical implementation, and operational integration. These characteristics necessitate a specialized framework that addresses the technical and organizational aspects of ML initiatives.

TABLE I
THE MACHINE LEARNING CANVAS

| Questions | Models | Dimensions | Canvas Element(s) |
|---|---|---|---|
| What value is created? | BMC | Value Proposition | Value Proposition |
| | MLC | Strategy | Task, Data, Target |
| How is the value created? | BMC | Value Architecture | Key Activities |
| | MLC | Process | Data Collection, Data Preparation, Algorithm Selection, Training & Evaluation, Performance Optimization, Risk Optimization |
| What is needed to create the value? | BMC | Value Infrastructure | Key Resources, Key Partners, Customer relationship, Customer segments, Channels |
| | MLC | Ecosystem | Deployment Infrastructure, Development Infrastructure, Production Integration |
| What drives the business impact of the created value? | BMC | Value Finance | Cost structure, Revenue Structure |
| | MLC | Support | Project Management & Governance Organizational Leadership |

## III. Methodology

To examine the determinants of success for an ML project and develop a theoretically based model, we mapped the BMC dimensions onto the corresponding ML constructs. These are strategy, process, ecosystem, and support. Next, we identified independent measures for each dimension from literature.

The dependent variable is IT project success, which is measured using the Iron Triangle framework, consisting of performance goals, business objectives, and stakeholder satisfaction [17]. Although data scientists frequently focus on technical performance measures, such as AUC scores or model accuracy, the Iron Triangle uses a language that business stakeholders understand: time, cost, and quality. Despite their technical complexity, ML projects are fundamentally business investments that must deliver value within constraints. A model that achieves 99% accuracy becomes irrelevant if it is delivered six months late and exceeds the budget by double. The Iron Triangle captures this reality by measuring what stakeholders care about: whether the project delivers its promised functionality on time and within budget.

TABLE II
CONSTRUCT MEASURES FOR IT PROJECT SUCCESS

| Item | The ML projects, you are involved, are typically perceived as successful, if … |
|---|---|
| S1 PERF | the ML model meets the performance goals. |
| S2 BO | the ML project results meet the business objectives. |
| S3 STAK | all stakeholders are satisfied with the ML project. |

The independent variables consist of different dimensions of the ML Canvas, which are measured on a 5-point Likert scale ranging from "strongly agree" to "strongly disagree." The SEM approach enables examination of how the four ML Canvas dimensions - Strategy, Process, Ecosystem, and Support - work together to influence project success. This methodology acknowledges that these dimensions are not directly observable, but rather latent constructs measured through multiple survey items.

### A. Strategy Dimension

The ML Strategy dimension translates business objectives into machine learning problems [18]. First, the task definition element requires mapping business problems to the appropriate learning paradigms to identify tasks suitable for machine learning [19]. This includes 1) distinguishing between prediction, exploration, and optimization problems, 2) assessing data availability, and 3) defining explainability requirements for the outcome [20].

Before starting any machine learning project, data specification is a necessary task. It specifies the data sources and modalities and distinguishes between structured and unstructured data and how to store and process them [21]. Quality considerations now extend beyond traditional database integrity to include representativeness, temporal stability, and potential biases that could compromise the model's validity [22]. Recent work emphasizes that data cascades, or compounding errors from poor data quality, are a primary cause of ML project failures [23].

The target definition connects technical performance metrics to business value creation. This addresses the hidden technical debt in machine learning systems [7]. Quantitative metrics should encompass not only accuracy or F1 scores, but also computational efficiency, latency constraints, and robustness to distribution shifts [24]. Qualitative considerations, such as interpretability requirements driven by regulatory frameworks like the GDPR and constraints that prevent discriminatory outcomes, have gained equal importance [25][26]. The emergence of responsible AI frameworks requires explicit consideration of transparency, accountability, and auditability throughout the ML lifecycle [27].

These conceptual elements operationalize the strategy dimension by turning it into measurable items for our empirical analysis.

TABLE III
CONSTRUCT MEASURES FOR THE STRATEGY DIMENSION

| Item | Strategy Dimension |
| --- | --- |
| ST1_TARGET | The success measures of the ML project are clearly defined. |
| ST2_TASK | The tasks in the ML project are clearly defined. |
| ST3_DATA | The requirements of the ML project, such as data and model constraints (explainability, space, inference time, …) are clearly defined. |

*B. Process Dimension*

The ML Process dimension uses an adapted CRISP-DM framework to transform raw data into deployable models. This framework acknowledges the iterative and non-deterministic nature of machine learning development [28].

Extraction-transform-load (ETL) processes are required for data collection to address big data's "5 V's": volume, variety, velocity, veracity, and value [29]. Volume encompasses storage and governance architectures that balance accessibility and security, as well as distributed systems that ensure scalability without sacrificing consistency [30]. Data lakes must accommodate both structured and unstructured data while maintaining traceability and version control. Variety requires integration pipelines that reconcile heterogeneity, such as relational schemas, unstructured text, and temporal streams, through semantic mapping and schema matching [31]. Robust encoding strategies are required for this: categorical variables need one-hot, target, or embedding approaches, which affect interpretability and dimensionality; text requires vectorization and embedding; and temporal features need cyclical encoding for seasonal patterns. Velocity introduces temporal complexity, where batch and stream processing coexist with different constraints on feature computation and model updating [32]. Standardization and normalization impact model convergence and offer different trade-offs in terms of outlier sensitivity and information preservation. Quality assessment encompasses ML-specific considerations, such as label noise tolerance, feature drift detection, and representation sufficiency [33]. Multidimensional quality - including completeness, consistency, timeliness, and relevance - requires composite metrics that reflect downstream performance rather than isolated characteristics [34].

Data preparation consumes up to 80% of project effort [35]. The cleansing process must balance information preservation with noise reduction. Outlier removal, for example, may eliminate rare but legitimate patterns that are crucial for predicting rare events [36]. Modern approaches use weak supervision and programmatic labeling to expand the reach of human expertise. These approaches encode domain knowledge as labeling functions rather than manual annotations [35]. Feature engineering is an important interface between domain knowledge and statistical learning. Practitioners must balance the automatic feature learning capabilities of automated machine learning (AutoML) or deep learning architectures with the interpretability and efficiency of handcrafted features using statistical methods or visual exploration of feature importance [37]. The feature selection and transformation pipeline embody increasing levels of abstraction. Selection methods (filter, wrapper, and embedded) optimize existing feature subsets. Extraction techniques (e.g., dimensionality reduction) discover latent representations. Transformations (e.g., normalization, discretization, and encoding) ensure algorithmic compatibility [38].

Algorithm selection is more than just finding a good algorithm that hits the performance targets. Formalized as the "No Free Lunch" theorem: no single algorithm dominates across all problem domains [39]. This fundamental constraint requires a portfolio approach, in which the choice of algorithms reflects the intersection of the characteristics of the problem (e.g., data distribution, noise levels, and feature types), operational constraints (e.g., latency, memory, and explainability), and organizational capabilities (e.g., expertise and maintenance capacity).

The rise of neural architecture search and AutoML has shifted the challenge from selecting among fixed architectures to navigating large configuration spaces, albeit sometimes at the cost of interpretability [40]. Transfer learning introduces a paradigm shift from learning from scratch to adapting knowledge. In this approach, pre-trained representations from large-scale datasets serve as initialization points, reducing sample complexity for subsequent tasks [41]. Ensemble methods operationalize the "wisdom of crowds" principle by achieving robustness through diversity via bagging's variance reduction, boosting's bias reduction, or stacking's hierarchical combination [42].

The training and evaluation processes must balance multiple competing objectives, including statistical validity, computational efficiency, and business relevance. The strategy for splitting the training data includes random partitioning, as well as temporal splits for time-series data, stratified sampling for imbalanced classes, and grouped splits for hierarchical data structures. Each of these strategies preserves different aspects of the data-generating process. Designing the objective function is an important step in translating business goals into mathematical optimization. Surrogate losses approximate true business metrics while maintaining differentiability and computational tractability [43]. Modern evaluation frameworks emphasize robustness across multiple dimensions, such as performance stability across data subgroups (slice-based evaluation), degradation under distribution shift (stress testing), and consistency across random seeds (variance analysis).

Performance optimization operates within the constraint triangle of model capacity, training data, and computational resources. Improvements in one area often require sacrifices in the others. Error analysis has evolved from merely inspecting a confusion matrix to using sophisticated frameworks such as error analysis loops and counterfactual debugging. These frameworks systematically probe model failures to identify fixable patterns [44]. The bias-variance decomposition theory provides a foundation for optimization strategies, such as addressing high bias by increasing model capacity or feature engineering and combating high variance by using regularization, data augmentation, or ensemble methods [45]. Hyperparameter optimization has advanced from grid search to Bayesian optimization methods that model the response surface. Recent multi-fidelity optimization advances have dramatically reduced computational costs [46].

Risk assessments in ML projects focus on specific failure modes that are unique to learning approaches [47]. These failure modes include data and algorithm bias, as well as model vulnerability and security. Data bias can manifest through historical prejudices encoded in the training data, biases in the selection of the data, and biases in the definition of the features. Each of these requires different mitigation strategies, ranging from reweighting to causal debiasing [48]. Algorithmic bias emerges from the inductive biases inherent in learning algorithms, such as the preference for simple hypotheses in regularized models, locality bias in nearest neighbor methods, and hierarchical bias in deep networks. These biases can amplify or counteract data biases in complex ways [49][50]. Adversarial vulnerabilities expose ML systems' susceptibility to malicious inputs; imperceptible perturbations can cause catastrophic misclassifications. This calls for defensive strategies ranging from adversarial training to certified robustness [51]. The intersection of these risks creates compound vulnerabilities, such as biased training data enabling targeted adversarial attacks or algorithmic preferences creating predictable failure modes. This requires holistic risk frameworks that consider interactions rather than isolated threats [52].

The process dimension is operationalized by key machine learning activities. We distinguish between preprocessing activities, such as data collection, cleaning, and feature engineering, and postprocessing activities, such as algorithm selection, training, evaluation, optimization, and risk assessment. We recognize that these phases require different skill sets and organizational competencies.

TABLE IV
CONSTRUCT MEASURES FOR THE PROCESS DIMENSION

| Item | Process Dimension: Preprocessing |
|---|---|
| PRE1_COLL | Gathering and consolidating the data is a straightforward task. |
| PRE2_CLEAN | Cleaning and preparing the data is a straightforward task. |
| PRE3_FEAT | Identifying the right features from the data is a straightforward task. |
| **Item** | **Process Dimension: Postprocessing** |
| POST1_RISK | Assessing the potential risk of the ML model is a straightforward process. |
| POST2_TRAIN | Training ML algorithms is a straightforward process. |
| POST3_OPTI | Optimizing a ML model is a straightforward process. |
| POST4_SEL | Selecting and testing different ML algorithms is a straightforward task. |
| POST5_EVAL | Evaluating a ML model is a straightforward task. |

### C. Ecosystem Dimension

The ML Ecosystem dimension represents the sociotechnical infrastructure that transforms ML experiments into production systems. Although ML algorithms comprise only 5% of real-world ML systems, the supporting infrastructure determines whether models provide value or remain costly experiments [53]. This ecosystem must reconcile competing demands: the exploratory nature of ML development, which requires flexibility and rapid iteration, and production requirements, which demand reliability, reproducibility, and governance.

The infrastructure architecture for ML projects has evolved from ad hoc assemblies to sophisticated platforms. The foundational layer integrates various data sources, including operational databases, data warehouses, streaming platforms, and application programming interfaces [54]. The compute layer's specialized hardware, ranging from GPUs to tensor-optimized TPUs, poses challenges for workload scheduling. Security layers address ML-specific concerns, such as differential privacy for training data, defenses against model extraction, and regulatory audit trails [55]. Although cloud platforms offer elastic scaling, preconfigured services, and access to cutting-edge hardware, data gravity, regulatory constraints, and latency requirements require hybrid architectures or even on-premises resources [56][57]. Cost optimization requires nuanced strategies, such as using spot instances for training, and reserved capacity for inference. Multi-cloud strategies are a way to avoid vendor lock-in and balance risk mitigation with operational complexity [58].

The ecosystem of development tools for ML requires integration across statistical computing environments, software engineering toolchains, and domain-specific platforms. Programming language choices embody fundamental trade-offs. For example, there is a trade-off between Python's ecosystem dominance and ease of use and the performance advantages of compiled languages. This often leads to polyglot architectures where Python serves as an orchestration layer [59]. ML frameworks themselves represent different philosophies: TensorFlow's production-first approach with static graphs; PyTorch's research-friendly, dynamic computation; and JAX's functional programming paradigm, which enables novel optimization strategies. Experiment tracking and versioning tools address the reproducibility crisis in ML, capturing not just code, but also hyperparameters, random seeds, and the computational environment [60]. Collaboration platforms bridge the gap between the notebook-based exploration favored by data scientists and the IDE-based development preferred by engineers while enabling peer review [61]. Monitoring and debugging tools face unique ML challenges. For example, detecting data drift, model degradation, and fairness violations requires statistical sophistication that goes beyond traditional application performance monitoring [62].

Production integration is the "almost last mile" problem in the ML project, where promising prototypes often fail to deliver value due to engineering challenges. Technical integration requires careful API design to handle the inherent uncertainty of ML predictions, versioning strategies to enable gradual rollout and rollback, and containerization approaches [63]. Scalability challenges manifest differently throughout the ML lifecycle. Training scalability is achieved through data and model parallelism. Serving scalability is achieved through batching and caching strategies. Feature computation scalability is achieved through incremental processing [64]. Model monitoring involves tracking key metrics and incorporating behavioral drift detection via statistical tests, adversarial input detection via uncertainty quantification, and fairness monitoring across demographic groups [65]. The democratization of ML via low-code platforms and AutoML services promises to increase accessibility while raising concerns about the trade-off between abstraction and control and the potential for creating "black box" infrastructure that mirrors the complexity of the models themselves [66].

TABLE IV
CONSTRUCT MEASURES FOR THE ECOSYSTEM DIMENSION

| Item | Ecosystem Dimension |
|---|---|
| ECO1_DEV | Development and collaboration during the ML project are a straightforward task. |
| ECO2_INT | Integrating the ML model in a production environment is a straightforward task. |
| ECO3_INFRA | Working with the provided IT infrastructure is a straightforward task. |

### D. Support Dimension

The support dimension addresses the "organizational learning" imperative in AI adoption - the recognition that technical excellence alone cannot guarantee ML project success without corresponding organizational capabilities and governance structures [67]. It encompasses managerial and leadership practices that bridge the "deployment gap" between ML potential and realized business value. This gap claims up to 87% of data science projects before production deployment [68].

Managing projects for ML initiatives requires finding a balance between the exploratory nature of data science with organizational demands for predictability and control. ML projects require hybrid teams that combine domain expertise, statistical knowledge, engineering capabilities, and business acumen [69]. To accommodate ML's inherent unpredictability, development methodologies must adapt agile principles, leading to approaches like "ML Sprint," which embeds experimentation cycles within broader delivery frameworks [70].

For ML project success, organizational leadership demands transformation across multiple dimensions of corporate capability. Stakeholder engagement must address the "AI trust deficit," which is the intersection of public skepticism about AI's fairness and transparency and employee fears about automation and job displacement. This requires communication strategies that emphasize augmentation over replacement and transparency over secrecy [71]. Performance measurement must also capture business value. This requires "AI transformation playbooks," which define success through operational improvements, customer satisfaction, and strategic positioning rather than model accuracy and quantitative performance alone [72]. Organizations must develop "AI fluency," or the organizational capacity to identify ML opportunities, manage ML projects, and integrate ML outputs into decision-making processes [73].

The intersection of project management and organizational leadership is evident in governance structures that balance innovation with responsibility. ML governance frameworks must address unique challenges that are not present in traditional IT governance. These challenges include model lifecycle management, versioning and deprecation policies, fairness and bias monitoring across protected attributes, explainability requirements for high-stakes decisions, and continuous monitoring for performance degradation [74]. Regulatory frameworks such as the EU's AI Act introduce external governance requirements. These requirements include risk assessments, transparency measures, and human oversight for high-risk AI applications. This transforms compliance from a technical checklist into a strategic imperative [75].

These business support elements reduce to three fundamental indicators of organizational commitment: whether ML projects receive company-wide recognition (SP1_WELL), strategic prioritization (SP2_PRIO), and adequate financial resources (SP3_FIN). These observable markers reflect the degree to which an organization has internalized and acted upon the complex support requirements described.

TABLE V
CONSTRUCT MEASURES FOR THE SUPPORT DIMENSION

| Item | Ecosystem Dimension |
|---|---|
| SP1_WELL | ML projects are well recognized across the company |
| SP2_PRIO | ML projects have a high priority in the company. |
| SP3_FIN | The financial budget for the ML project is sufficient. |

## IV. MODELLING

The following model's structure reflects the fundamental principle that machine learning projects exhibit unique characteristics requiring specific organizational capabilities and sequential dependencies [7][14]. The model's structure addresses the fundamental tension in ML projects between exploration and exploitation [76]. Organizational support enables exploration through resource provision, strategy channels this exploration toward specific objectives, processes exploit chosen approaches through technical implementation, and ecosystems institutionalize successful patterns. This progression from exploration to exploitation through increasing specification explains why alternative causal structures would violate the inherent logic of ML development.

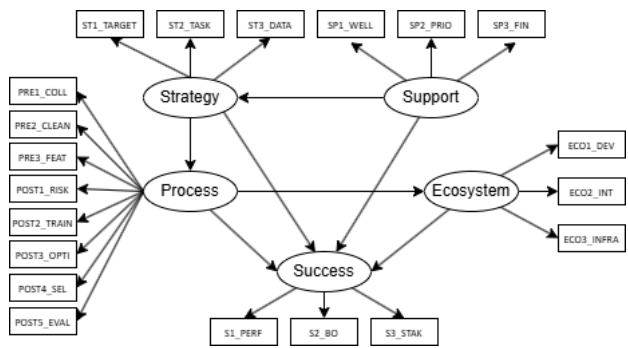

Fig. 1. Structured Equation Model based on the dimensions of the Machine Learning Canvas, reflecting the causal dependencies between the constructs.

The positioning of organizational support as the initial causal driver aligns with resource-based view theory, which posits that organizational resources and capabilities determine strategic options [77]. This causal relationship is not merely correlational; experimental evidence shows that variations in organizational support directly influence the quality and scope of data-driven strategic initiatives [78]. The absence of direct paths from support to downstream elements reflects "the analytics gap" - resources alone cannot generate value without strategic translation [79].

The central role of strategy as a mediator between support and process implementation finds strong theoretical grounding in the strategic alignment literature. IT strategy serves as a critical bridge between organizational capabilities and technical execution [80]. For ML projects, strategic clarity - operationalized through task definition, data specification, and target setting - determines the entire trajectory of technical development [81].

The causal flow from strategy to process reflects the inherent logic of ML development, where strategic decisions preceding technical implementation. ML projects following a strategy-first approach achieve significantly higher success rates than those beginning with technical experimentation [82].

In ML contexts, deployment infrastructure needs to emerge from, rather than determine, technical process decisions [8]. This causal direction is empirically supported by Google's ML engineering practices, where teams first establish data pipelines and model architectures before designing deployment systems [14].

## V. RESULTS

To test whether this structured approach leads to successful ML projects, we surveyed data scientists. The survey underwent two rounds of pilot testing before being distributed to 3,000 data scientists, yielding 161 responses (a 5% response rate). After cleaning the data, we deemed 150 responses suitable for analysis, where the data scientists use AI coding assistant on a daily base (1-2 hours on average). Most of the respondents were employed professionals (67%), followed by students (20%), freelancers (11%), and business owners (9%). Forty percent worked in large enterprises with more than 500 employees, while the rest represented companies with fewer than 20 employees (15%), 20–50 employees (6%), 50–100 employees (8%), or 100–500 employees (13%). Nineteen percent selected N/A.

The participants came from a variety of sectors, including fintech, education, artificial intelligence (AI), online retail, financial services, analytics consulting, and telecommunications. Of these participants, only 14% reported ML projects with real business impact. Twenty-two percent leaned toward business impact, 25% were neutral, and 39% characterized their work as experimental (23% leaned toward experimental and 16% were purely pilot projects). This distribution reflects the range of perspectives on the ML adoption spectrum, from experimental initiatives to production deployments.

Prior to conducting the structural equation modeling analysis, we verified that all fundamental assumptions were satisfied. First, multivariate normality was assessed using Mahalanobis distance with chi-square distribution testing. No outliers were detected as all p-values exceeded the 0.001 threshold, confirming the multivariate normality assumption. Multicollinearity diagnostics revealed no concerns, with all Variance Inflation Factor values below 10 and tolerance values exceeding 0.1, indicating absence of collinearity among predictor variables.

TABLE VI
TOLERANCE VALUES AND VARIANCE INFLATION FACTOR

| Model | Unstand. Coeff. B | Std. Error | Stand. Coeff. Beta | t | Sig. | Tole-rance | VIF |
|---|---|---|---|---|---|---|---|
| (Constant) | 90.72 | 22.792 |  | 3.98 | 0 |  |  |
| S1_PERF | -6.477 | 7.888 | -0.094 | -0.821 | 0.413 | 0.531 | 1.882 |
| S2_BO | 4.612 | 7.156 | 0.075 | 0.645 | 0.52 | 0.521 | 1.921 |
| S3_STAK | 4.811 | 6.458 | 0.08 | 0.745 | 0.458 | 0.607 | 1.646 |
| ST1_TARGET | -0.46 | 4.789 | -0.011 | -0.096 | 0.924 | 0.553 | 1.809 |
| ST2_TASK | -2.115 | 5.499 | -0.051 | -0.385 | 0.701 | 0.403 | 2.482 |
| ST3_DATA | -6.998 | 5.373 | -0.161 | -1.303 | 0.195 | 0.454 | 2.205 |
| SP1_WELL | 6.084 | 4.922 | 0.142 | 1.236 | 0.219 | 0.527 | 1.899 |
| SP2_PRIO | -5.894 | 5.005 | -0.136 | -1.178 | 0.241 | 0.526 | 1.902 |
| SP3_FIN | 2.997 | 4.917 | 0.065 | 0.61 | 0.543 | 0.619 | 1.615 |
| PRE1_COLL | -9.476 | 6.303 | -0.223 | -1.503 | 0.135 | 0.315 | 3.171 |
| PRE2_CLEAN | 15.531 | 6.474 | 0.362 | 2.399 | 0.018 | 0.306 | 3.268 |
| PRE3_FEAT | -3.156 | 5.694 | -0.069 | -0.554 | 0.58 | 0.455 | 2.195 |
| POST1_RISK | 4.513 | 6.38 | 0.096 | 0.707 | 0.481 | 0.38 | 2.629 |
| POST2_TRAIN | -4.781 | 5.7 | -0.102 | -0.839 | 0.403 | 0.467 | 2.143 |
| POST3_OPTI | 1.383 | 5.876 | 0.031 | 0.235 | 0.814 | 0.396 | 2.527 |
| POST4_SEL | -1.418 | 5.641 | -0.028 | -0.251 | 0.802 | 0.553 | 1.807 |
| POST5_EVAL | -0.576 | 7.132 | -0.012 | -0.081 | 0.936 | 0.309 | 3.241 |
| ECO1_DEV | -3.199 | 6.169 | -0.069 | -0.519 | 0.605 | 0.388 | 2.575 |
| ECO2_INT | -3.197 | 5.218 | -0.077 | -0.613 | 0.541 | 0.439 | 2.28 |
| ECO3_INFRA | 3.348 | 6.148 | 0.077 | 0.545 | 0.587 | 0.349 | 2.864 |

The tolerance values (all above 0.30) and Variance Inflation Factor (VIF) scores (all below 5.0) indicate acceptable levels of multicollinearity, confirming that the SEM assumptions regarding independent variables' linear independence are met.

Heteroskedasticity was examined through residual plots with loess smoothing, which displayed a straight-line pattern, confirming homoscedasticity. The correlation matrix demonstrated positive definiteness with a determinant greater than zero, satisfying this critical assumption for SEM estimation.

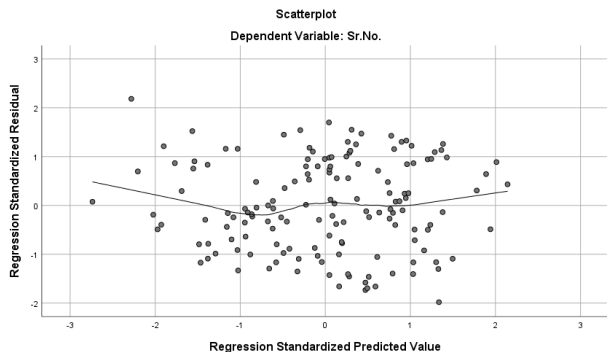

Fig. 2. Heteroskedasticity / homoscedasticity analysis

Initial factor loadings revealed that the financial support item (SP3_FIN) exhibited excessive cross-loadings, compromising discriminant validity. Following established procedures, this item was removed from further analysis, resulting in improved model specification. Reliability analysis demonstrated strong internal consistency across all constructs, with Cronbach's alpha values exceeding the 0.7 threshold for all dimensions. The refined model is as follows:

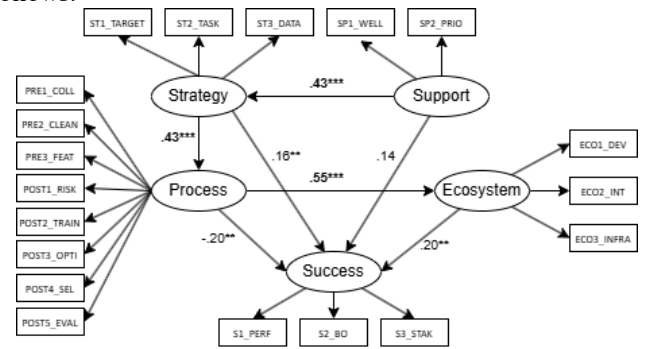

Fig. 3. Refined model of success determinants for a machine learning project, without the SP3_FIN measure. *** indicates significance at 1% level and ** at 5% significance level.

TABLE VII
REGRESSION WEIGHTS OF THE REFINED MODEL

| Dependent | Independent | Estimate | S.E. | C.R. | PLabel |
|---|---|---|---|---|---|
| Strategy | Support | 0.432 | 0.12 | 3.601 | *** |
| Process | Strategy | 0.428 | 0.086 | 5.006 | *** |
| Ecosystem | Process | 0.547 | 0.092 | 5.93 | *** |
| Success | Strategy | 0.163 | 0.081 | 2.021 | 0.043 |
| Success | Support | 0.137 | 0.076 | 1.809 | 0.071 |
| Success | Ecosystem | 0.202 | 0.08 | 2.524 | 0.012 |
| Success | Process | -0.199 | 0.09 | -2.217 | 0.027 |
| ST1_TARGET | Strategy | 0.783 | 0.101 | 7.769 | *** |
| ST2_TASK | Strategy | 1 |  |  |  |
| ST3_DATEN | Strategy | 0.898 | 0.102 | 8.778 | *** |
| SP1_WELL | Support | 1 |  |  |  |
| SP2_PRIO | Support | 0.829 | 0.185 | 4.471 | *** |
| ECO3_INFRA | Ecosystem | 1 |  |  |  |
| ECO2_INT | Ecosystem | 0.876 | 0.098 | 8.972 | *** |
| ECO1_DEV | Ecosystem | 0.697 | 0.089 | 7.823 | *** |
| POST4_SEL | Process | 0.899 | 0.093 | 9.68 | *** |
| POST3_OPTI | Process | 0.838 | 0.088 | 9.49 | *** |
| POST2_TRAIN | Process | 1 |  |  |  |
| POST1_RISK | Process | 0.982 | 0.087 | 11.296 | *** |
| PRE3_FEAT | Process | 0.863 | 0.087 | 9.909 | *** |
| PRE2_CLEAN | Process | 0.939 | 0.093 | 10.088 | *** |
| PRE1_COLL | Process | 0.909 | 0.096 | 9.47 | *** |
| S3_STAK | Success | 0.834 | 0.13 | 6.399 | *** |
| S2_BO | Success | 1 | | | |
| POST5_EVAL | Process | 0.919 | 0.098 | 9.341 | *** |
| S1_PERF | Success | 0.902 | 0.13 | 6.958 | *** |

Confirmatory Factor Analysis (CFA) established strong convergent validity across all constructs. All Average Variance Extracted (AVE) values range from 0.58 to 0.74, comfortably surpassing the recommended threshold of 0.50, thereby confirming strong convergent validity. Composite Reliability (CR) values are consistently high (0.84 to 0.92), exceeding the commonly accepted benchmark of 0.70, thus indicating excellent internal consistency. The factor loadings (square root of CR), ranging between 0.92 and 0.96, demonstrate robust associations between each construct and its observed indicators. Additionally, error variances, calculated as (1 - CR), remain low (0.08 to 0.16), further underscoring the high reliability and precision of the measurement model.

TABLE VIII
RESULTS OF THE CFA

| Measure | Success | Process | Ecosystem | Support | Strategy |
|---|---|---|---|---|---|
| AVE | 0.63 | 0.58 | 0.64 | 0.74 | 0.63 |
| CR | 0.84 | 0.92 | 0.84 | 0.85 | 0.84 |
| Factor Loading | 0.92 | 0.96 | 0.92 | 0.92 | 0.92 |
| Error variance | 0.16 | 0.08 | 0.16 | 0.15 | 0.16 |

Discriminant validity was confirmed using the Fornell-Larcker criterion. All AVE values for each construct exceeded the squared correlations with other constructs. The highest squared correlation (0.26) occurred between Ecosystem and Process, which remained below both constructs' AVE values, confirming that each construct captures unique variance not shared with other constructs.

Results of the discriminant validity analysis (DVA) show the highest squared correlation (0.26) occurred between Ecosystem and Process, which remained below both constructs' AVE values, confirming that each construct captures unique variance not shared with other constructs.

TABLE IX
RESULTS OF THE DVA

| | | | Factor Correl. | Correl. squared | AVE1 | AVE2 |
|---|---|---|---|---|---|---|
| Strategy | <--> | Process | 0.467 | 0.22 | 0.63 | 0.58 |
| Support | <--> | Process | 0.184 | 0.03 | 0.74 | 0.58 |
| Ecosystem | <--> | Process | 0.507 | 0.26 | 0.64 | 0.58 |
| Process | <--> | Success | 0.037 | 0.00 | 0.58 | 0.63 |
| Strategy | <--> | Support | 0.427 | 0.18 | 0.63 | 0.74 |
| Strategy | <--> | Ecosystem | 0.315 | 0.10 | 0.63 | 0.64 |
| Strategy | <--> | Success | 0.31 | 0.10 | 0.63 | 0.63 |
| Support | <--> | Ecosystem | 0.145 | 0.02 | 0.74 | 0.64 |
| Support | <--> | Success | 0.314 | 0.10 | 0.74 | 0.63 |
| Ecosystem | <--> | Success | 0.26 | 0.07 | 0.64 | 0.63 |

The evaluation of model fit assesses how well the theoretical model aligns with the observed data. the Normed Fit Index (NFI) of 0.867 and Relative Fit Index (RFI) of 0.841, slightly below the recommended threshold (≥ 0.90), suggest minor improvements could be made regarding the relative comparison to a baseline model. However, strong incremental fit measures—such as the Incremental Fit Index (IFI = 0.960), Tucker-Lewis Index (TLI = 0.951), and Comparative Fit Index (CFI = 0.959)—demonstrate that the hypothesized structural model provides substantial improvements over a baseline independence model. Moreover, the Root Mean Square Error of Approximation (RMSEA = 0.050) indicates excellent absolute fit, clearly within acceptable boundaries (≤ 0.06), supported by a favorable PCLOSE value (0.480). Overall, these results indicate that the structural relationships postulated in the SEM model represent the data effectively. Overall, the SEM analysis reveals good model performance, with particularly robust indicators from IFI, TLI, CFI, and RMSEA

TABLE X
RESULT OF THE MODEL FIT ANALYSIS.

| Model | NFI (Delta1) | RFI (rho1) | IFI (Delta2) | TLI (rho2) | CFI |
|---|---|---|---|---|---|
| Default model | 0.867 | 0.841 | 0.96 | 0.951 | 0.959 |
| Saturated model | 1 | 1 | 1 | 1 | 1 |
| Independence model | 0 | 0 | 0 | 0 | 0 |

| Model | RMSEA | LO 90 | HI 90 | PCLOSE |
|---|---|---|---|---|
| Default model | 0.05 | 0.031 | 0.067 | 0.48 |
| Independence model | 0.226 | 0.216 | 0.237 | 0 |

All hypothesized mediator relationships demonstrated statistical significance with z-values exceeding 1.96. Strategy mediates the relationship between Support and Process, while Process mediates between Strategy and Ecosystem. These serial mediation effects reveal the cascading nature of ML project success factors, where organizational support initiates a chain of effects through strategy clarity, structured processes, and implementation quality, ultimately determining project success.

## VI. DISCUSSION

### A. Support -> Strategy (highly significant)

Organizational support significantly influences strategy formulation, aligning with established information systems literature on the importance of organizational backing for IT project success. The significant relationship between organizational support and strategy formulation takes on new dimensions in the era of LLM-assisted development. As GitHub Copilot studies demonstrate, developers using AI coding assistants can achieve 55% faster task completion, but this acceleration requires clear organizational guidelines and strategic frameworks [5]. The support dimension becomes critical for establishing AI usage policies, ethical guidelines, and resource allocation for LLM tools. This aligns with organizational support structures determine whether AI tools amplify productivity or create confusion [83]. Organizations must now provide support not just for traditional ML projects but for meta-level decisions about when and how to leverage LLMs in the development process.

### B. Strategy -> Process (highly significant)

The path from strategy to structured preprocessing and postprocessing reveals a paradox in LLM-assisted development. While LLMs can generate preprocessing code rapidly, our findings suggest that strategic clarity remains essential for effective implementation [84]. This supports that developers using Copilot still require clear problem formulation and data understanding - tasks that LLMs cannot fully automate [85].

### C. Process -> Ecosystem (highly significant)

While LLMs excel at generating boilerplate preprocessing code, AI-generated code often lacks the robustness necessary for production systems [86]. Our findings suggest that structured processes remain crucial even when leveraging LLMs, supporting that LLM-assisted development requires new forms of quality assurance and testing [87]. The ecosystem perspective becomes essential as teams must integrate LLM-generated components with existing infrastructure, monitoring systems, and governance frameworks

### D. Ecosystem -> Success (significant)

The relationship between ecosystem and project success aligns with recent findings on LLM-assisted development outcomes. While LLMs accelerate individual coding tasks, project success depends on systematic integration practices [88]. The ecosystem approach becomes critical when integrating models into specific organizational contexts [89]. Success requires not just using these tools but building comprehensive ecosystems that handle versioning, monitoring, and continuous improvement of both human-written and AI-generated code.

### E. Strategy -> Success (significant)

The direct effect of strategy on success, beyond its mediated effects, supports that in AI-augmented environments, strategic clarity becomes more, not less, important [90]. The direct path suggests that strategy provides value in addition to operational efficiency - it shapes how teams conceptualize problems, evaluate LLM suggestions, and make architectural decisions that no current AI system can fully automate.

### F. Process -> Success (significant)

The negative coefficient likely represents a suppressor effect, where Process only contributes positively to Success through its influence on Ecosystem implementation [91]. This statistical phenomenon occurs when the direct effect has an opposite sign from the total effect, indicating that Process activities consume resources and create complexity that, without proper deployment infrastructure, actually hinder success.

In summary, these findings reveal that LLMs fundamentally alter the "how" but not the "why" of ML project success. While coding assistants can accelerate technical implementation, our model demonstrates that success still flows through organizational support, strategic clarity, and structured processes. This supports the "AI pair programming" paradigm, where LLMs augment rather than replace human expertise [92].

## VII. LIMITATIONS

Several limitations of this study should be considered when interpreting the results. First, while the sample size of 150 respondents is adequate for SEM analysis, it may limit the generalizability of the findings to different organizational contexts and industries. The 5% response rate raises concerns about non-response bias because those who completed the survey may be data scientists with specific characteristics or experiences. Second, despite the theoretical grounding of our model, the cross-sectional design prevents causal inference; longitudinal studies would better capture the dynamic nature of ML project evolution. Third, the self-reported nature of the data may introduce common method bias, though our multi-factor structure and discriminant validity tests mitigate this concern to some extent. Fourth, our focus on data scientists may not represent all roles in machine learning project contexts. Finally, removing the financial support item (SP3_FIN) due to cross-loadings suggests that our operationalization of the support dimension may not capture all relevant organizational factors, especially those related to resource allocation.

## VIII. SUMMARY

This research addresses the persistent challenge of ML project failure by developing and validating a comprehensive framework that acknowledges the technical and organizational dimensions of success. The MLC adapts business model thinking to the specific context of ML development. It provides a structured approach that remains relevant despite the acceleration enabled by LLM coding assistants. Our empirical findings show that LLMs may revolutionize the "how" of ML development through rapid code generation. However, the fundamental success factors—strategic alignment, structured processes, robust ecosystems, and organizational support—remain unchanged. Our SEM analysis revealed cascading relationships that offer practical insights: Organizations cannot adopt AI coding tools and expect improved outcomes. Instead, they must build comprehensive capabilities across all four dimensions. Thus, the MLC framework serves as a diagnostic tool for identifying weaknesses in current ML initiatives and as a planning instrument for future projects, helping organizations navigate the intersection of human expertise and AI assistance in modern ML development.

> REPLACE THIS LINE WITH YOUR PAPER IDENTIFICATION NUMBER (DOUBLE-CLICK HERE TO EDIT) <    11[41] F. Zhuang, Z. Qi, K. Duan, D. Xi, Y. Zhu, H. Zhu, H. Xiong, and Q. He, "A comprehensive survey on transfer learning," *Proceedings of the IEEE*, vol. 109, no. 1, pp. 43-76, 2020.

[42] Z. H. Zhou, *Ensemble methods: Foundations and algorithms*. Boca Raton, FL: CRC Press, 2021.

[43] R. C. Williamson and A. K. Menon, "Fairness risk measures," in *Proceedings of the 36th International Conference on Machine Learning*, pp. 6786-6797, 2019.

[44] T. Wu, M. T. Ribeiro, J. Heer, and D. Weld, "Errudite: Scalable, reproducible, and testable error analysis," in *Proceedings of the 57th Annual Meeting of the Association for Computational Linguistics*, pp. 747-763, 2019.

[45] M. Belkin, D. Hsu, S. Ma, and S. Mandal, "Reconciling modern machine-learning practice and the classical bias–variance trade-off," *Proceedings of the National Academy of Sciences*, vol. 116, no. 32, pp. 15849-15854, 2019.

[46] L. Li, K. Jamieson, A. Rostamizadeh, E. Gonina, M. Hardt, B. Recht, and A. Talwalkar, "A system for massively parallel hyperparameter tuning," *Proceedings of Machine Learning and Systems*, vol. 2, pp. 230-246, 2020.

[47] R. Ashmore, R. Calinescu, and C. Paterson, "Assuring the machine learning lifecycle: Desiderata, methods, and challenges," *ACM Computing Surveys*, vol. 54, no. 5, pp. 1-39, 2022.

[48] H. Suresh and J. Guttag, "A framework for understanding sources of harm throughout the machine learning life cycle," in *Equity and Access in Algorithms, Mechanisms, and Optimization*, pp. 1-9, 2021.

[49] T. M. Mitchell, "The need for biases in learning generalizations," Technical Report CBM-TR-117, Rutgers University, 1980.

[50] R. Baeza-Yates, "Bias on the web," *Communications of the ACM*, vol. 61, no. 6, pp. 54-61, 2018.

[51] A. Madry, A. Makelov, L. Schmidt, D. Tsipras, and A. Vladu, "Towards deep learning models resistant to adversarial attacks," in *International Conference on Learning Representations*, 2018.

[52] R. S. S. Kumar, M. Nyström, J. Lambert, A. Marshall, M. Goertzel, A. Comissoneru, M. Swann, and S. Xia, "Adversarial machine learning-industry perspectives," in *2020 IEEE Security and Privacy Workshops*, pp. 69-75, 2020.

[53] N. Polyzotis, S. Roy, S. E. Whang, and M. Zinkevich, "Data lifecycle challenges in production machine learning: A survey," *ACM SIGMOD Record*, vol. 47, no. 2, pp. 17-28, 2018.

[54] A. Ratner et al., "MLSys: The new frontier of machine learning systems," arXiv preprint arXiv:1904.03257, 2019.

[55] T. Li, A. K. Sahu, A. Talwalkar, and V. Smith, "Federated learning: Challenges, methods, and future directions," *IEEE Signal Processing Magazine*, vol. 37, no. 3, pp. 50-60, 2020.

[56] E. Jonas, J. Schleier-Smith, V. Sreekanti, C. C. Tsai, A. Khandelwal, Q. Pu, V. Shankar, J. Carreira, K. Krauth, N. Yadwadkar, J. E. Gonzalez, R. A. Popa, I. Stoica, and D. A. Patterson, "Cloud programming simplified: A Berkeley view on serverless computing," arXiv preprint arXiv:1902.03383, 2019.

[57] M. Satyanarayanan, "The emergence of edge computing," *Computer*, vol. 50, no. 1, pp. 30-39, 2017.

[58] R. Bhardwaj, A. Nambiar, and D. Dutta, "A study of machine learning in cloud computing," *IEEE Transactions on Cloud Computing*, vol. 10, no. 1, pp. 166-183, 2022.

[59] S. Raschka, J. Patterson, and C. Nolet, "Machine learning in Python: Main developments and technology trends in data science, machine learning, and artificial intelligence," *Information*, vol. 11, no. 4, 193, 2020.

[60] M. Zaharia, R. S. Xin, P. Wendell, T. Das, M. Armbrust, A. Dave, X. Meng, J. Rosen, S. Venkataraman, M. J. Franklin, A. Ghodsi, J. Gonzalez, S. Shenker, and I. Stoica, "Apache Spark: A unified engine for big data processing," *Communications of the ACM*, vol. 59, no. 11, pp. 56-65, 2016.

[61] A. Rule, A. Tabard, and J. D. Hollan, "Exploration and explanation in computational notebooks," in *Proceedings of the 2018 CHI Conference on Human Factors in Computing Systems*, pp. 1-12, 2018.

[62] E. Breck, N. Polyzotis, S. Roy, S. E. Whang, and M. Zinkevich, "Data validation for machine learning," *Proceedings of Machine Learning and Systems*, vol. 1, pp. 334-347, 2019.

[63] D. Kang, D. Raghavan, P. Bailis, and M. Zaharia, "Model assertions for monitoring and improving ML models," *Proceedings of Machine Learning and Systems*, vol. 2, pp. 481-496, 2020.

[64] T. Ben-Nun and T. Hoefler, "Demystifying parallel and distributed deep learning: An in-depth concurrency analysis," *ACM Computing Surveys*, vol. 52, no. 4, pp. 1-43, 2019.

[65] J. Gama, I. Žliobaitė, A. Bifet, M. Pechenizkiy, and A. Bouchachia, "A survey on concept drift adaptation," *ACM Computing Surveys*, vol. 46, no. 4, pp. 1-37, 2014.

[66] D. Xin, L. Ma, J. Liu, S. Macke, S. Song, and A. Parameswaran, "Accelerating human-in-the-loop machine learning: Challenges and opportunities," in *Proceedings of the Fourth Workshop on Data Management for End-to-End Machine Learning*, pp. 1-8, 2021.

[67] S. Ransbotham, S. Khodabandeh, R. Fehling, B. LaFountain, and D. Kiron, "Winning with AI," *MIT Sloan Management Review*, vol. 61, no. 1, pp. 17-19, 2019.

[68] VentureBeat, "Why do 87% of data science projects never make it into production?" July. 2019. [Online]. https://venturebeat.com/ai/why-do-87-of-data-science-projects-never-make-it-into-production/2019.

[69] T. H. Davenport and D. J. Patil, "Data scientist: The sexiest job of the 21st century," *Harvard Business Review*, vol. 90, no. 10, pp. 70-76, 2012.

[70] Google Cloud, "Practitioners guide to MLOps: A framework for continuous delivery and automation of machine learning," Google Cloud Architecture Framework, 2021.

[71] E. Glikson and A. W. Woolley, "Human trust in artificial intelligence: Review of empirical research," *Academy of Management Annals*, vol. 14, no. 2, pp. 627-660, 2020.

[72] A. Ng, "The AI transformation playbook," Landing AI, 2021.

[73] J. K. U. Brock and F. von Wangenheim, "Demystifying AI: What digital transformation leaders can teach you about realistic artificial intelligence," *California Management Review*, vol. 61, no. 4, pp. 110-134, 2019.

[74] P. Saleiro, B. Kuester, L. Hinkson, J. London, A. Stevens, A. Anisfeld, K. T. Rodolfa, and R. Ghani, "Aequitas: A bias and fairness audit toolkit," arXiv preprint arXiv:1811.05577, 2018.

[75] European Commission, "Proposal for a regulation laying down harmonised rules on artificial intelligence (Artificial Intelligence Act)," COM(2021) 206 final, 2021.

[76] J. G. March, "Exploration and exploitation in organizational learning," *Organization Science*, vol. 2, no. 1, pp. 71-87, 1991.

[77] J. Barney, "Firm resources and sustained competitive advantage," *Journal of Management*, vol. 17, no. 1, pp. 99-120, 1991.

[78] E. Brynjolfsson and K. McElheran, "The rapid adoption of data-driven decision-making," *American Economic Review*, vol. 106, no. 5, pp. 133-139, 2016.

[79] T. H. Davenport and J. G. Harris, *Competing on analytics: The new science of winning*. Boston, MA: Harvard Business Press, 2007.

[80] J. C. Henderson and N. Venkatraman, "Strategic alignment: Leveraging information technology for transforming organizations," *IBM Systems Journal*, vol. 32, no. 1, pp. 4-16, 1993.

[81] T. Fountaine, B. McCarthy, and T. Saleh, "Building the AI-powered organization," *Harvard Business Review*, July-August 2019.

[82] K. Wagstaff, "Machine learning that matters," in *Proceedings of the 29th International Conference on Machine Learning*, pp. 529-536, 2012.

[83] E. Brynjolfsson, D. Li, and L. Raymond, "Generative AI at work," National Bureau of Economic Research Working Paper, no. w31161, 2023.

[84] M. Chen et al., "Evaluating large language models trained on code," arXiv preprint arXiv:2107.03374, 2021.

[85] A. Sarkar, A. D. Gordon, C. Negreanu, C. Poelitz, S. S. Ragavan, and B. Zorn, "What is it like to program with artificial intelligence?" arXiv preprint arXiv:2208.06213, 2021.

[86] J. D. Weisz, M. Muller, S. Houde, J. Richards, S. I. Ross, F. Martinez, M. Agarwal, and K. Talamadupula, "Perfection not required? Human-AI partnerships in code translation," in *26th International Conference on Intelligent User Interfaces*, pp. 402-412, 2021.

[87] T. Zhang, V. Kishore, F. Wu, K. Q. Weinberger, and Y. Artzi, "BERTScore: Evaluating text generation with BERT," arXiv preprint arXiv:1904.09675, 2023.

[88] P. Vaithilingam, T. Zhang, and E. L. Glassman, "Expectation vs. experience: Evaluating the usability of code generation tools powered by large language models," in *CHI Conference on Human Factors in Computing Systems Extended Abstracts*, pp. 1-7, 2022.

**Martin Prause:** As a scientist and entrepreneur, he combines approaches from economics, computer science, and behavioral economics to make decision-making processes - both at the individual and organizational level - transparent and comprehensible. He is the co-founder and CTO of a company that uses simulations to make corporate decision-making visible in order to optimize decisions and successfully integrate the resulting insights into business strategy.